\pdfoutput=1

\documentclass[3p,times]{elsarticle}

\usepackage{ecrc}

\usepackage[breaklinks]{hyperref}
\usepackage[utf8x]{inputenc}
\usepackage{subfig}
\usepackage{amsmath}
\usepackage{graphicx}

\volume{00}

\firstpage{1}

\journalname{Nuclear Physics A}

\runauth{T. Lappi and H. Mäntysaari}

\jid{nupha}

\jnltitlelogo{Nuclear Physics A}

\usepackage{amssymb}

\biboptions{comma,square,sort&compress}

\usepackage[figuresright]{rotating}

\newcommand{\der}{\mathrm{d}}
\newcommand{\rt}{{\mathbf{r}_T}}

\newcommand{\bt}{{\mathbf{b}_T}}
\newcommand{\st}{{\mathbf{s}_T}}

\newcommand{\pt}{{\mathbf{p}_T}}
\newcommand{\qt}{{\mathbf{q}_T}}
\newcommand{\kt}{{\mathbf{k}_T}}

\newcommand{\cf}{C_\mathrm{F}}

\newcommand{\nr}[1]{(\ref{#1})}

\newcommand{\qs}{Q_\mathrm{s}}

\newcommand{\qso}{Q_\mathrm{s0}}

\newcommand{\lqcd}{\Lambda_{\mathrm{QCD}}}
\newcommand{\as}{\alpha_{\mathrm{s}}}

\newcommand{\eq}{Eq.~}
\newcommand{\se}{Sec.~}

\newcommand{\Ncal}{{\mathcal{N}}}

\begin{document}



\hypersetup{pdfauthor={T. Lappi and H. M\"antysaari},pdftitle={Particle Production in Color Class Condensate: from electron-proton DIS to proton-nucleus collisions}}



\dochead{}


\author{T. Lappi}
\address{
Department of Physics, %
 P.O. Box 35, 40014 University of Jyv\"askyl\"a, Finland and \\
 Helsinki Institute of Physics, P.O. Box 64, 00014 University of Helsinki,
Finland
}

\author{H. M\"antysaari}
\address{
Department of Physics, %
 P.O. Box 35, 40014 University of Jyv\"askyl\"a, Finland
}

\title{
Particle Production in the Color Class Condensate: from electron-proton DIS to proton-nucleus collisions
}

\begin{abstract}

We study single inclusive hadron production in proton-proton and 
proton-nucleus collisions in the CGC framework. The parameters in the 
calculation are obtained by fitting  electron-proton deep inelastic
 scattering data. The obtained dipole-proton amplitude is 
generalized to dipole-nucleus scattering without any additional nuclear parameters
other than the Woods-Saxon distribution.
We show that it is possible to use an initial condition without an
anomalous dimension and still obtain a good description of the HERA inclusive
cross section and LHC single particle production measurements.
We argue that one must consistently 
use the proton transverse area as measured by a high virtuality probe in 
DIS also for the single inclusive cross section in proton-proton and 
proton-nucleus collisions, and obtain a nuclear modification factor
$R_{pA}$ that at midrapidity approaches unity at large momenta and
at all energies. 

\end{abstract}

\maketitle

\section{Introduction}
Non-linear phenomena, such as gluon recombination, manifest themselves in
strongly interacting systems at high energy. A convenient way to describe
these effects is provided by the Color Glass Condensate effective field theory.
Because the gluon densities scale as $A^{1/3}$, these phenomena are 
enhanced when the target is changed from a proton to a heavy nucleus.
The p+Pb run at the LHC allows us to probe the non-linearly behaving QCD matter
in a kinematical region never explored so far.  


The structure of a hadron can be studied accurately in deep inelastic 
scattering (DIS) where a (virtual) photon scatters off the hadron. Precise
measurements of the proton structure at HERA have been a crucial test for the 
CGC, and recent analyses have confirmed that the CGC
description is consistent with all the available small-$x$ DIS data, see e.g.~ 
\cite{Albacete:2010sy}.
In addition to DIS, within the CGC framework it is possible to simultaneously describe also other  
high-energy hadronic interactions.
In addition to single inclusive particle production discussed in this work, these include, for example,
two-particle correlations~\cite{Lappi:2012nh,Albacete:2010pg,Dusling:2013oia} and diffractive
DIS~\cite{Kowalski:2006hc,Lappi:2013am}.
Comparing calculations fit to DIS data 
to particle production results for proton-proton and proton-nucleus collisions
at different energies provides a nontrivial test of the
universality of the CGC description, and makes 
predictions for future LHC pA measurements.

In this work we compute, as explained in more detail in Ref. \cite{Lappi:2013zma}, 
hadron production in proton-proton and proton-nucleus collisions consistently within the CGC
framework. As an input we use only the HERA data for the inclusive DIS cross section and standrad nuclear geometry, and the generalization to nuclei is done without any additional nuclear parameters.

\section{Electron-proton baseline}
\label{sec:ep}


\begin{figure}
\begin{minipage}[t]{0.48\linewidth}
\centering
\includegraphics[width=1.05\textwidth]{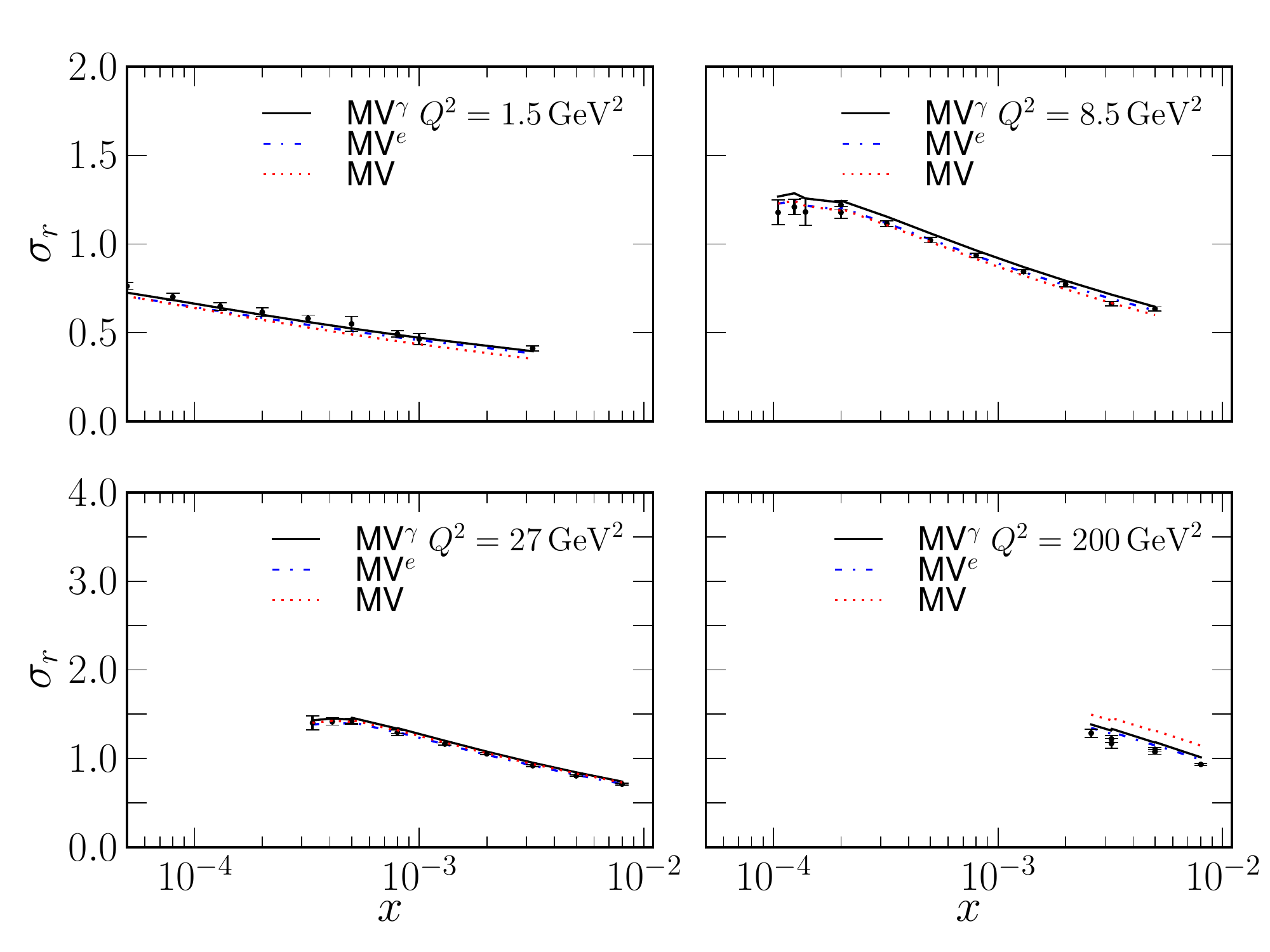} 
\caption{
Reduced cross section $\sigma_r$ computed using the MV$^\gamma$, MV$^e$ and MV model initial conditions 
for the dipole amplitude compared with combined HERA 
(H1 and ZEUS) data~\cite{Aaron:2009aa}.}
\label{fig:sigmar}
\end{minipage}
\hspace{0.5cm}
\begin{minipage}[t]{0.48\linewidth}
\centering
\includegraphics[width=1.05\textwidth]{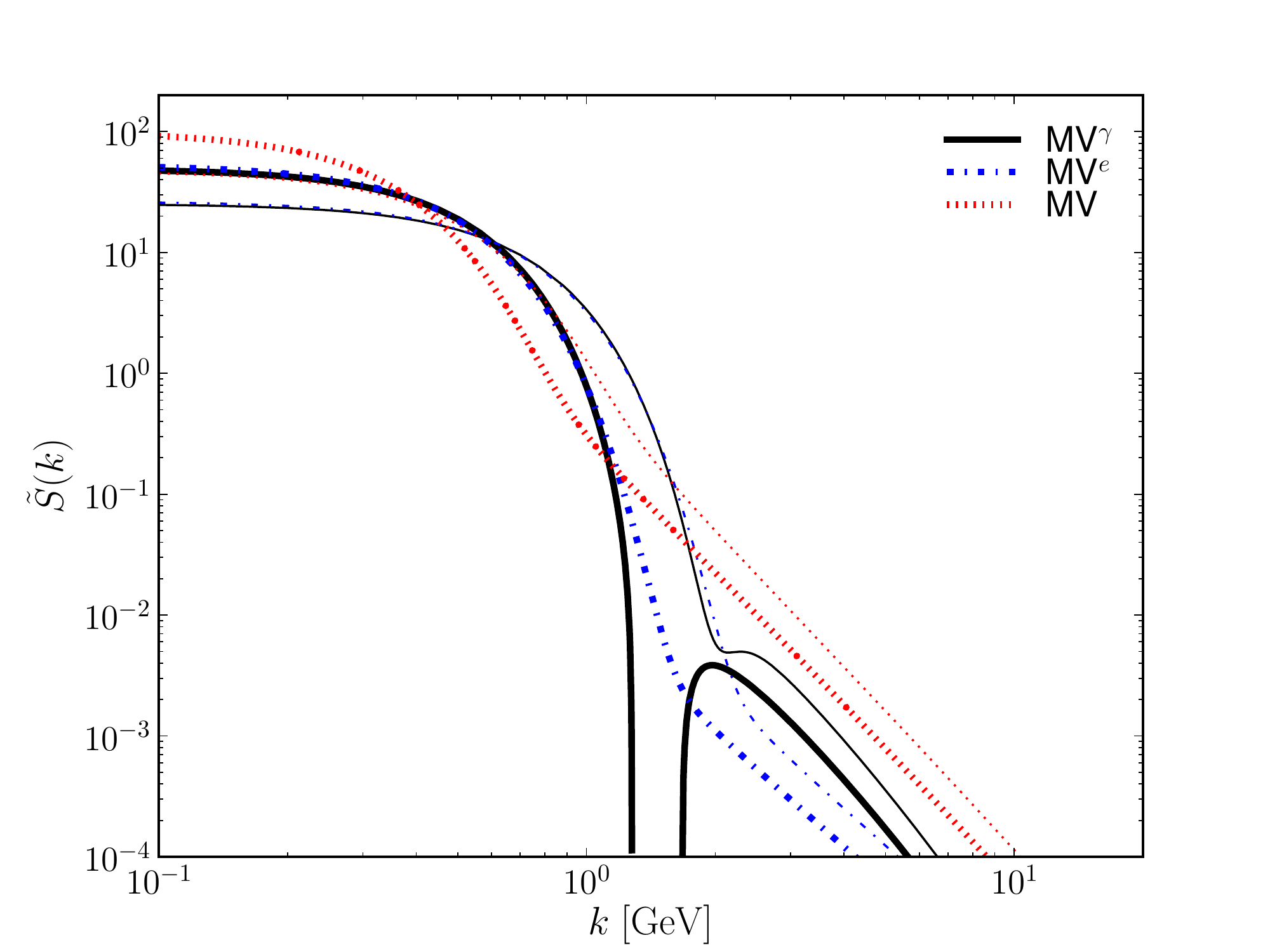} 
\caption{
Two dimensional Fourier transform of $S(r)=1-\Ncal(r)$ in fundamental (thick lines) and adjoint (thin lines) representations for MV$^\gamma$, MV$^e$ and MV models.
}\label{fig:ft-s}
\end{minipage}
\end{figure}



Deep inelastic scattering provides a precision measurement of the
proton structure. The H1 and ZEUS collaborations have measured the proton
structure functions $F_2$ and $F_L$ and published very precise combined
results for the reduced cross section $\sigma_r$~\cite{Aaron:2009aa}, 
which is a linear combination of the proton structure functions:
\begin{equation}
	\sigma_r(y,x,Q^2) = F_2(x,Q^2) - \frac{y^2}{1+(1-y)^2} F_L(x,Q^2).
\end{equation}
Here $y=Q^2/(sx)$ and $\sqrt{s}$ is the center of mass energy. The structure
functions are related to the virtual photon-proton cross sections 
$\sigma_{T,L}^{\gamma^*p}$ for transverse (T) and longitudinal (L) photons that can, in the Color Glass Condensate framework, be computed as
%
\begin{equation}\label{eq:sigmagammastar}
	\sigma_{T,L}^{\gamma^*p}(x,Q^2) = 2\sum_f \int \der z \int \der^2 \bt |\Psi_{T,L}^{\gamma^* \to f\bar f}|^2 \Ncal(\bt, \rt, x).
\end{equation}
Here $\Psi_{T,L}^{\gamma^* \to f\bar f}$ is the photon light cone wave 
function describing how the photon fluctuates to a quark-antiquark pair,
computed from light cone QED~\cite{Kovchegov:2012mbw}. 
The QCD dynamics is 
inside the function $\Ncal(\bt,\rt,x)$ which is the imaginary part of the scattering
amplitude for the process where a dipole scatters off the color field of a hadron
with impact parameter $\bt$. It can not be computed perturbatively, but its
energy (or equivalently Bjorken $x$) dependence satisfies the BK 
equation, for which we use the running
coupling corrections derived in Ref.~\cite{Balitsky:2006wa}. 

We assume here that the impact parameter dependence of the proton 
factorizes and one replaces
\begin{equation}
	2\int \der^2 \bt \to \sigma_0.
\end{equation}
Note that with the usual convention adopted here the factor 2 from the optical theorem 
is absorbed into the constant $\sigma_0$, thus 
the transverse area of the proton is now $\sigma_0/2$. The proton area could in principle be obtained from diffractive vector meson production, see discussion in Ref \cite{Lappi:2013zma}. In this work, however, we consider $\sigma_0$ to be a fit parameter.

\begin{table}
\begin{center}
\begin{tabular}{|l||r|r|r|r|r|r|r|}
\hline
Model & $\chi^2/\text{d.o.f}$ &  $\qso^2$ [GeV$^2$] & $\qs^2$ [GeV$^2$] & $\gamma$ & $C^2$ & $e_c$ & $\sigma_0/2$ [mb] \\
\hline\hline
MV & 2.76 & 0.104 & 0.139  & 1 & 14.5 & 1 & 18.81 \\
MV$^\gamma$ & 1.17 & 0.165 & 0.245 & 1.135 & 6.35 & 1 & 16.45 \\
MV$^e$ & 1.15 & 0.060 & 0.238  & 1 & 7.2 & 18.9 & 16.36 \\
\hline
\end{tabular}
\caption{Parameters from fits to HERA reduced cross section data at $x<10^{-2}$ and $Q^2<50\,\mathrm{GeV}^2$ for different initial conditions. Also the corresponding initial saturation scales $\qs^2$ defined via equation $\Ncal(r^2=2/\qs^2)=1-e^{-1/2}$ are shown. The parameters for the MV$^\gamma$ initial condition are obtained by the AAMQS collaboration \cite{Albacete:2010sy}.
}
\label{tab:params}
\end{center}
\end{table}

As a non-perturbative input one needs also the dipole-proton amplitude at the 
initial $x=x_0$, for which we use a modified
 McLerran-Venugopalan model~\cite{McLerran:1994ni}:
\begin{equation}
\label{eq:aamqs-n}
	\Ncal(\rt) = 1 - \exp \left[ -\frac{(\rt^2 \qso^2)^\gamma}{4} \ln \left(\frac{1}{|\rt| \lqcd}+e_c \cdot e\right)\right],
\end{equation}
where we have generalized the AAMQS~\cite{Albacete:2010sy} form by also
allowing the constant inside the logarithm, which plays a role of an infrared cutoff, to be different from $e$. The other fit parameters are the
anomalous dimension $\gamma$ and the initial saturation scale $\qso^2$.

The BK equation with running coupling requires the strong coupling constant 
$\as$ as a function of the transverse separation $r=|\rt|$. In order to obtain 
a slow enough evolution to be compatible with the data (see discussion in 
Ref.~\cite{Kuokkanen:2011je}) we include, as in Ref. \cite{Albacete:2010sy}, 
an additional fit parameter $C^2$ such that (see also Ref. \cite{Lappi:2012vw} for a discussion of the numerical value of $C^2$)
\begin{equation}
	\as(r) = \frac{12\pi}{(33 - 2N_f) \log \left(\frac{4C^2}{r^2\lqcd^2} \right)},
\end{equation}
with $\lqcd$ fixed to the value $0.241$ GeV.

The unknown parameters are obtained by performing a fit to small-$x$ DIS
data. The first parametrization considered in this work, denoted by 
MV$^\gamma$, is obtained by setting $e_c \equiv 1$ in \eq \eqref{eq:aamqs-n}
but keeping the anomalous dimension $\gamma$ as a fit parameter.
This is fitted by the AAMQS collaboration in 
Ref.~\cite{Albacete:2010sy}, resulting in a very good 
$\chi^2/\text{d.o.f}\approx 1.17$. The obtained values for the parameters are listed in Table 
\ref{tab:params}.

We fit the two other initial conditions for the dipole amplitude to the HERA 
reduced cross section data. First, we consider a parametrization, denoted here
by MV$^e$, where we
do not include anomalous dimension ($\gamma \equiv 1$) but let $e_c$ be a 
free fit parameter. The second model studied for comparison is the MV model
without modifications, where $\gamma \equiv 1$ and $e_c \equiv 1$. 
The fit quality obtained using the MV$^e$ initial condition
is essentially as good as for the
MV$^\gamma$ model, with the best fit giving $\chi^2/\text{d.o.f}\approx 1.15$.
The fit parameters are listed in Table \ref{tab:params}.
Viewed in momentum space
this parametrization provides a smoother interpolation 
between the small-$k$ saturation region (where it resembles the Gaussian
GBW form) and the power law at behavior high $k$. 
This is demonstrated
in Fig. \ref{fig:ft-s}, where we show the Fourier-trasform of 
$S(\rt) = 1-\Ncal(\rt)$, which is proportional to the ``dipole'' gluon distribution.


The parameters obtained for the pure MV model, where
$\gamma=1$ and $e_c=1$, are listed in Table \ref{tab:params}. 
The fit quality is not as good as with the modified MV model, 
($\chi^2/\text{d.o.f}\sim 2.8$), but as one can see from Fig. \ref{fig:sigmar}, the 
description of the small-$x$ DIS data is still reasonable.


\section{Single inclusive hadron production in CGC}
\label{sec:sinc}
The gluon spectrum in heavy ion collisions can be obtained
by solving the classical Yang-Mills equations of
motion for the color fields. For $k_T\gtrsim \qs$ 
it has been shown numerically~\cite{Blaizot:2010kh}
that this solution is well approximated by the 
following $k_T$-factorized formula ~\cite{Kovchegov:2001sc}
\begin{equation}
\label{eq:ktfact-bdep}
\frac{\der \sigma}{\der y \der^2 \kt \der^2 \bt} = \frac{2 \as}{\cf \kt^2 } \int \der^2 \qt \der^2 \st \frac{\varphi_p(\qt,\st)}{\qt^2} 
	 \frac{\varphi_p(\kt-\qt,\bt-\st)}{(\kt-\qt)^2}.
\end{equation}
Here $\varphi_p$ is the dipole unintegrated gluon
distribution (UGD) of the 
proton~\cite{Kharzeev:2003wz,Blaizot:2004wu,Dominguez:2011wm} 
and $\bt$ is the 
impact parameter. For the proton we assume that the impact parameter dependence 
factorizes and 
\begin{equation}
\label{eq:ugd}
	\varphi_p(\kt) = \int \der^2 \bt \varphi_p(\kt, \bt) 
	 = \frac{\cf \sigma_0/2}{8\pi^3 \as} \kt^4 \tilde S^p(\kt).
\end{equation}
Here $\tilde S^p(k)$ is the two dimensional Fourier transform of the 
dipole-proton scattering matrix $S^p(r)=1-\Ncal^p_A(r)$, where $\Ncal^p_A$ is the
dipole-proton scattering amplitude in adjoint representation: 
$\Ncal_A = 2\Ncal-\Ncal^2$.
For the proton DIS area $\sigma_0/2$ we use the value from the fits to DIS data, 
see \se \ref{sec:ep}. 

Let us now consider a proton-proton collision. The cross section is obtained by 
integrating \eq \eqref{eq:ktfact-bdep} over the impact parameter, which gives
\begin{equation}
	\frac{\der \sigma}{\der y \der^2 \kt} = \frac{(\sigma_0/2)^2}{(2\pi)^2} \frac{\cf}{2\pi^2 \kt^2 \as} \int \frac{\der^2 \qt}{(2\pi)^2} \qt^2 \tilde S^p(\qt)
	  (\kt-\qt)^2 \tilde S^p(\kt-\qt).
\end{equation}
The invariant yield is defined as the production cross section 
divided by the total inelastic cross section $\sigma_\text{inel}$ and  thus becomes
\begin{equation}
\label{eq:ktfact-pp}
	\frac{\der N}{\der y \der^2 \kt} = \frac{(\sigma_0/2)^2}{\sigma_\text{inel}} \frac{\cf}{8\pi^4  \kt^2 \as} \int \frac{\der^2 \qt}{(2\pi)^2} \qt^2 \tilde S^p(\qt) 
		(\kt-\qt)^2 \tilde S^p(\kt-\qt).
\end{equation}


Assuming that $|\kt|$ is much larger than the saturation scale of one of the
protons we obtain the hybrid formalism result
\begin{equation}
	\label{eq:pp-hybrid}
	\frac{\der N}{\der y \der^2 \kt} = \frac{\sigma_0/2}{\sigma_\text{inel}} \frac{1}{(2\pi)^2} xg(x,\kt^2) \tilde S^p(\kt),
\end{equation}
where
\begin{equation}
	\label{eq:xg}
	xg(x,\kt^2) = \int_0^{\kt^2} \frac{\der \qt^2}{\qt^2} \varphi_p(\qt)
\end{equation}
is the integrated gluon distribution function. This can then 
be replaced by the conventional parton distribution function, for which we can use the 
CTEQ LO~\cite{Pumplin:2002vw} pdf.

HERA measurements of diffractive vector meson 
electroproduction~\cite{Aaron:2009xp} indicate that the proton transverse area 
measured with a high virtuality probe is smaller than in soft interactions. 
In our case this shows up
as a large difference in the numerical values of 
$\sigma_0/2$ and $\sigma_\text{inel}$, and leads to an energy 
dependent factor
$\frac{\sigma_0/2}{\sigma_\text{inel}}\sim 0.2\dots 0.3$ in the 
particle yield~\nr{eq:pp-hybrid}, in contrast with the treatment  
often used in CGC calculations.
Physically this corresponds  to a two-component picture
of the transverse structure of the nucleon (see also
Ref.~\cite{Frankfurt:2010ea} for a very similar discussion).
The small-$x$ gluons responsible for semihard particle production
occupy a small area $\sim \sigma_0/2$ in the core of the nucleon. 
This core is surrounded by a nonperturbative edge that becomes larger
with $\sqrt{s}$, but only participates in soft interactions that contribute
to the large total inelastic cross section $\sigma_\text{inel}$.
We will in Sec.~\ref{sec:nuke} show that this separation between 
the two transverse areas brings much clarity to the extension of 
the calculation from protons to nuclei.

Now that also the normalization ($\sigma_0/2$) from HERA data is used in the 
calculation of the single inclusive spectrum, the result represents the  actual LO CGC 
prediction for also the normalization of the spectrum. As is often
the case for perturbative QCD, the LO result only agrees with data within a factor of 
$\sim 2$.  
We therefore  multiply the resulting spectrum with a 
``$K$-factor'' to bring it to the level of the experimental data. 
Now that the different areas $\sigma_0$ and $\sigma_\text{inel}$ are properly
included, this factor has a more conventional interpretation of 
the expected effect of NLO corrections on the result; although it depends
quite strongly on the fragmentation function.

In order to obtain a hadron spectrum from the parton spectrum we 
convolute the cross section with the DSS LO fragmentation 
function~\cite{deFlorian:2007aj} and,
when using the hybrid formalism, also add the light quark-initiated channel
to the gluonic one in \eq\nr{eq:pp-hybrid}.

\section{From proton to nucleus}
\label{sec:nuke}

Due to a lack of small-$x$ nuclear DIS data we can not perform a
 similar fit to nuclear targets than what is done with the proton. 
Instead we use the optical Glauber model to generalize our dipole-proton 
amplitude to dipole-nucleus scattering. 

First we observe that the total dipole (size $r$)-proton 
cross section reads
\begin{equation}
	\sigma_\text{dip}^p = \sigma_0 \Ncal^p(r).
\end{equation}
In the dilute limit of very small dipoles the dipole-nucleus cross section 
should be just an incoherent sum of dipole-nucleon cross sections, i.e.
$\sigma_\text{dip}^A = A\sigma_\text{dip}^p$.
 On the other hand for large 
dipoles we should have $\der \sigma_\text{dip}^A / \der^2 \bt 
\equiv 2 \Ncal^A(\rt,\bt) \leq 2$. These 
requirements are satisfied with an exponentiated dipole-nucleus scattering amplitude
\begin{equation}\label{eq:glaubersigmadipa}
\Ncal^A(\rt,\bt) = \left[ 1 - \exp\left( -\frac{AT_A(\bt)}{2} \sigma_\text{dip}^p \right) \right].
\end{equation}
This form is an average of the dipole cross section over the fluctuating
positions of the nucleons in the nucleus (see e.g. \cite{Kowalski:2007rw}), 
and thus incorporates in an analytical expression the fluctuations discussed 
e.g. in Ref.~\cite{Albacete:2012xq}.

Using the form \nr{eq:glaubersigmadipa} directly in computing particle production is, 
however problematic. Because the forward $S$-matrix element
 $S=1-\Ncal^A(\rt,\bt)$ approaches
a limiting value $\exp\left( -\frac{AT_A(\bt)}{2} \sigma_0 \right) 
\sim \exp\left(-A^{1/3}\right) $ and not exactly zero at large $r$, the dipole 
gluon distribution  develops unphysical oscillations as a function of $k$.
We therefore expand the proton-dipole cross section in  \eq\nr{eq:glaubersigmadipa} 
and use the approximation 
\begin{equation}
	\sigma_\text{dip}^p = \sigma_0 \Ncal^p(\rt) \approx \sigma_0 \frac{(\rt^2 \qso^2)^\gamma}{4} \ln \left(\frac{1}{|\rt|\lqcd}+e_c \cdot e\right)
\end{equation}
in the exponent of \eq\nr{eq:glaubersigmadipa}.
The dipole-nucleus amplitude is then obtained by solving the rcBK evolution 
equation with an initial condition
\begin{equation}\label{eq:ainitc}
	\Ncal^A(\rt,\bt) = 1 - \exp\left[ -A T_A(\bt) \frac{\sigma_0}{2} \frac{(\rt^2 \qso^2)^\gamma}{4}  \ln \left(\frac{1}{|\rt|\lqcd}+e_c \cdot e\right) \right].
\end{equation}
We emphasize that besides the Woods-Saxon nuclear density $T_A(\bt)$, all the
parameters in this expression result from the fit to HERA data. The ``optical Glauber'' initial condition \nr{eq:ainitc} also
brings to evidence the advantage of the MV$^e$ parametrization, which  
achieves a good fit to HERA data while imposing $\gamma=1$.
 In contrast to the MV$^\gamma$ fit, this functional form avoids the ambiguity
encountered in e.g. Ref.~\cite{Albacete:2012xq}
of whether the factor $AT_A(\bt) \sigma_0/2$ should be replaced by 
$(AT_A(\bt) \sigma_0/2)^\gamma$ to achieve a natural scaling of 
$\qs^2$  with the nuclear thickness.

The fully impact parameter dependent BK equation develops unphysical Coulomb
 tails which would need an additional screening mechanism at the confinement scale
(see e.g.~\cite{GolecBiernat:2003ym}). We 
therefore  solve the scattering amplitudes for each $\bt$ independently. 
Due to the rapid increase of the scattering amplitude at low densities 
(large $|\bt|$) this effectively causes the nucleus to grow rapidly on the 
edges at large energies. 
We do not consider this parametrization
to be reliable at large $|\bt|$, and when computing particle production at peripheral collisions we assume that $R_{pA}=1$ when the saturation scale is below the proton saturation scale, see dicussion in Ref. \cite{Lappi:2013am}.


When computing proton-nucleus cross sections we compute convolutions of the 
nuclear and the proton unintegrated gluon distributions. In terms of the 
dipole amplitudes the $k_T$ factorization formula now reads
\begin{equation}
	\frac{\der N(\bt)}{\der y \der^2 \kt} = \frac{\sigma_0/2}{(2\pi)^2} \frac{\cf}{2\pi^2 \kt^2 \as} \int \frac{\der^2 \qt }{(2\pi)^2} \qt^2 \tilde S^p(\qt) 
	(\kt-\qt)^2 \tilde S^A(\kt-\qt),
\end{equation}
where $\tilde S^p$ and $\tilde S^A$ are Fourier transforms of the dipole-proton 
and dipole-nucleus scattering matrices, respectively. Assuming moreover that 
the transverse momentum of the produced parton is much larger than the proton 
saturation scale we get the hybrid formalism result
\begin{equation}
	\frac{\der N(\bt)}{\der y \der^2 \kt} = \frac{1}{(2\pi)^2} xg(x,\kt^2) \tilde S^A(\kt).
\end{equation}
Notice that, in contrast to \eq\nr{eq:pp-hybrid}, in this case we do not get a factor
$(\sigma_0/2)/\sigma_\text{inel}$ in the yield. One can then analytically show that this parametrization yields 
$R_{pA}\to 1$ at midrapidity and at large transverse momenta~\cite{Lappi:2013am} at all $\sqrt{s}$, even as $\sigma_\text{inel}$ and thus $N_\text{bin}$
are changing with  $\sqrt{s}$ while the initial saturation scale $\qso$ is not.

In proton-nucleus collisions it is not possible to determine the impact 
parameter by measuring the total multiplicity as well as in heavy ion
collisions, due to the large multiplicity fluctuations for a fixed impact parameter.
 The first LHC proton-lead results are
divided into centrality classes based on the multiplicity or energy deposit in 
forward calorimeters. This is a difficult quantity to handle theoretically,
so we assume that we can obtain reasonable estimates for different centrality 
classes by using a standard optical Glauber model. 
This difference should be kept in mind when comparing our calculations with the
corresponding LHC results.

\section{Results}
\label{sec:results}

\begin{figure}
\begin{minipage}[t]{0.48\linewidth}
\centering
\includegraphics[width=1.05\textwidth]{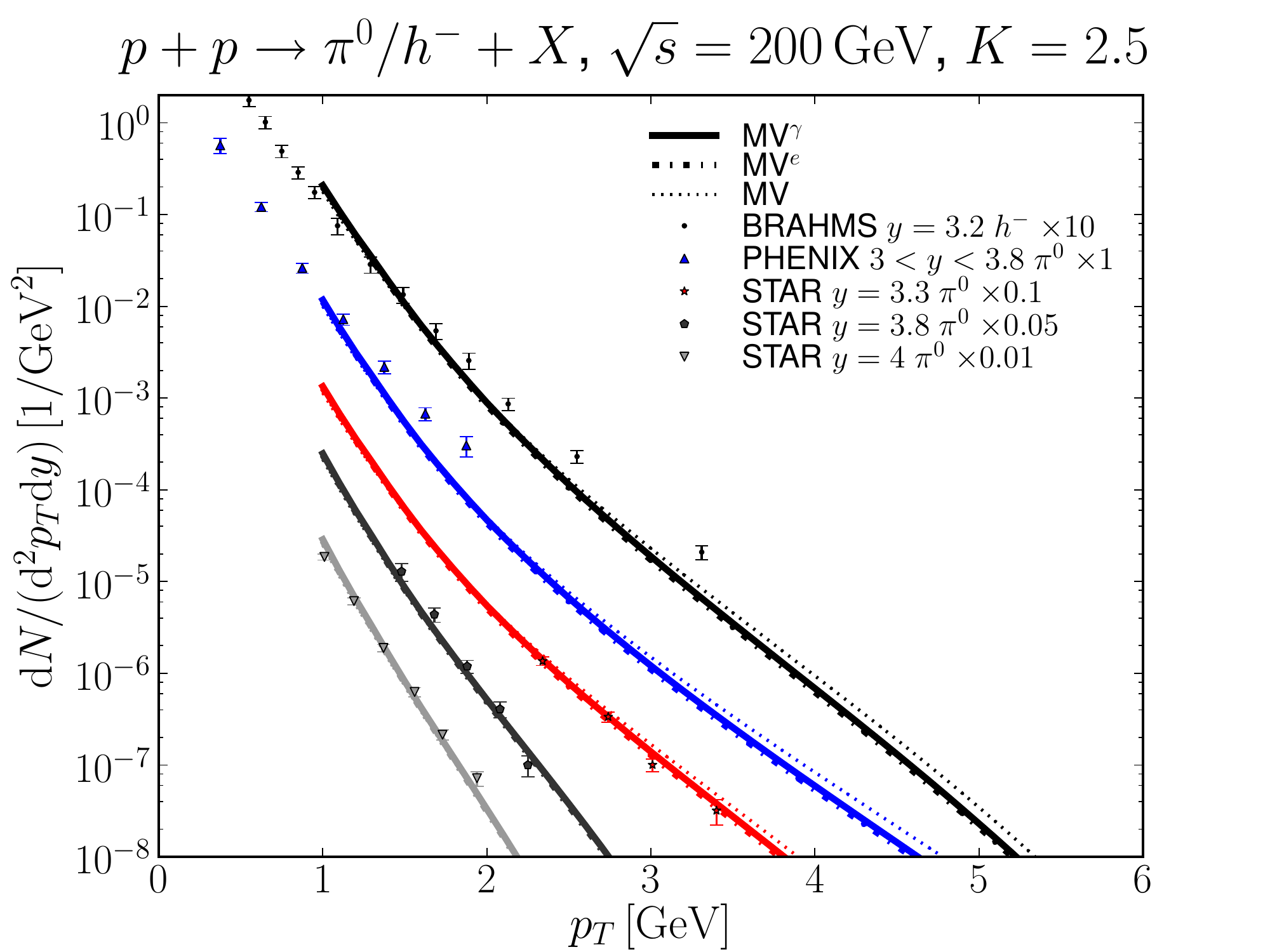} 
\caption{
Single inclusive $\pi^0$ and negative hadron production computed using MV, MV$^e$ and MV$^\gamma$ initial conditions compared with RHIC data from STAR~\cite{Adams:2006uz}, PHENIX~\cite{Adare:2011sc} and BRAHMS~\cite{Arsene:2004ux} collaborations. \\
}\label{fig:rhic_pp_yield}
\end{minipage}
\hspace{0.5cm}
\begin{minipage}[t]{0.48\linewidth}
\centering
\includegraphics[width=1.05\textwidth]{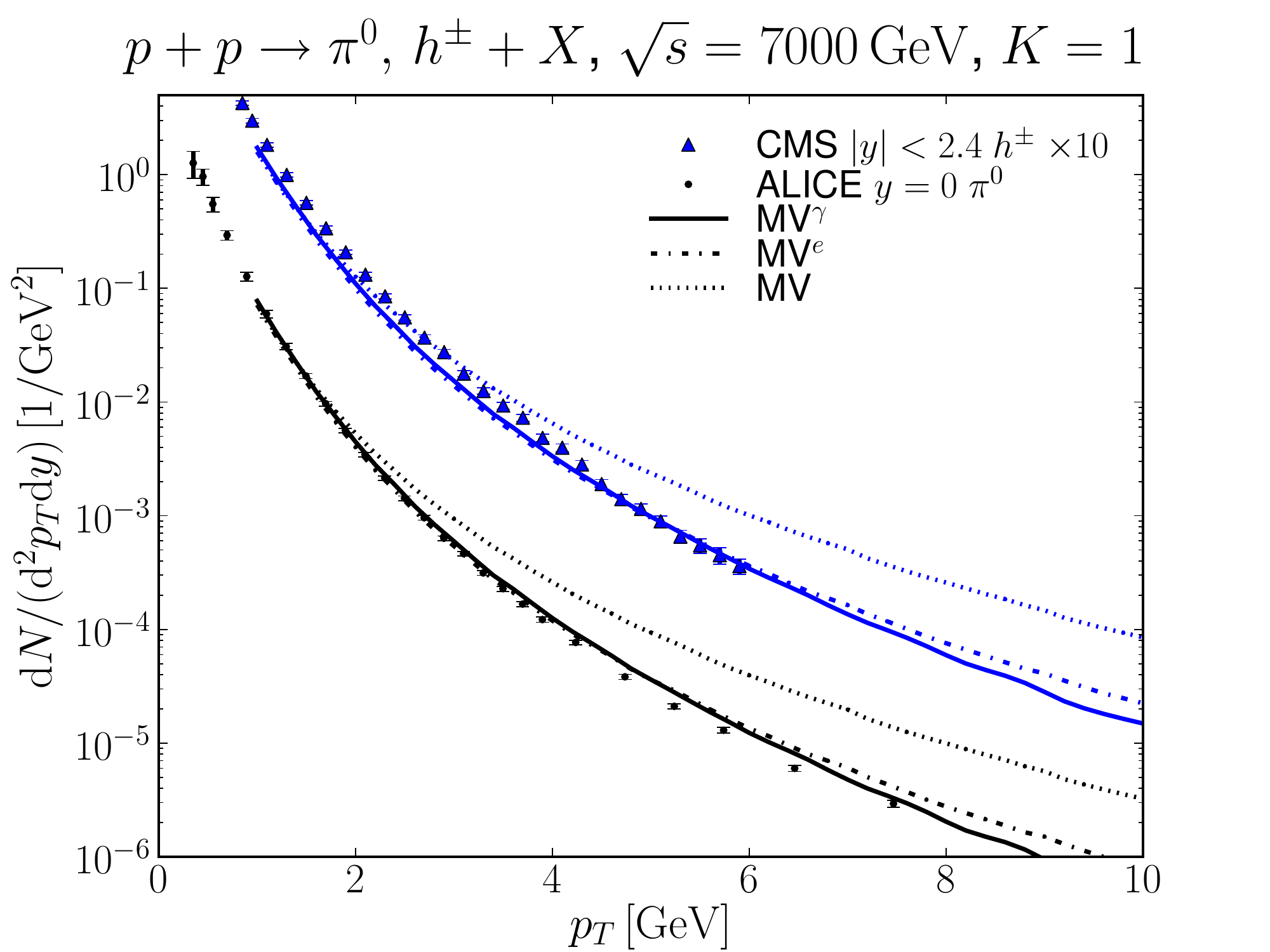} 
\caption{
Single inclusive $\pi^0$ production computed using MV, MV$^\gamma$ and MV$^e$ initial conditions at $\sqrt{s}=7000$ GeV compared with ALICE $\pi^0$ \cite{Abelev:2012cn} and CMS charged hadron data~\cite{Khachatryan:2010us}.\\
}\label{fig:lhc_pp_yield}
\end{minipage}
\end{figure}

\begin{figure}
\begin{minipage}[t]{0.48\linewidth}
\centering
\includegraphics[width=1.05\textwidth]{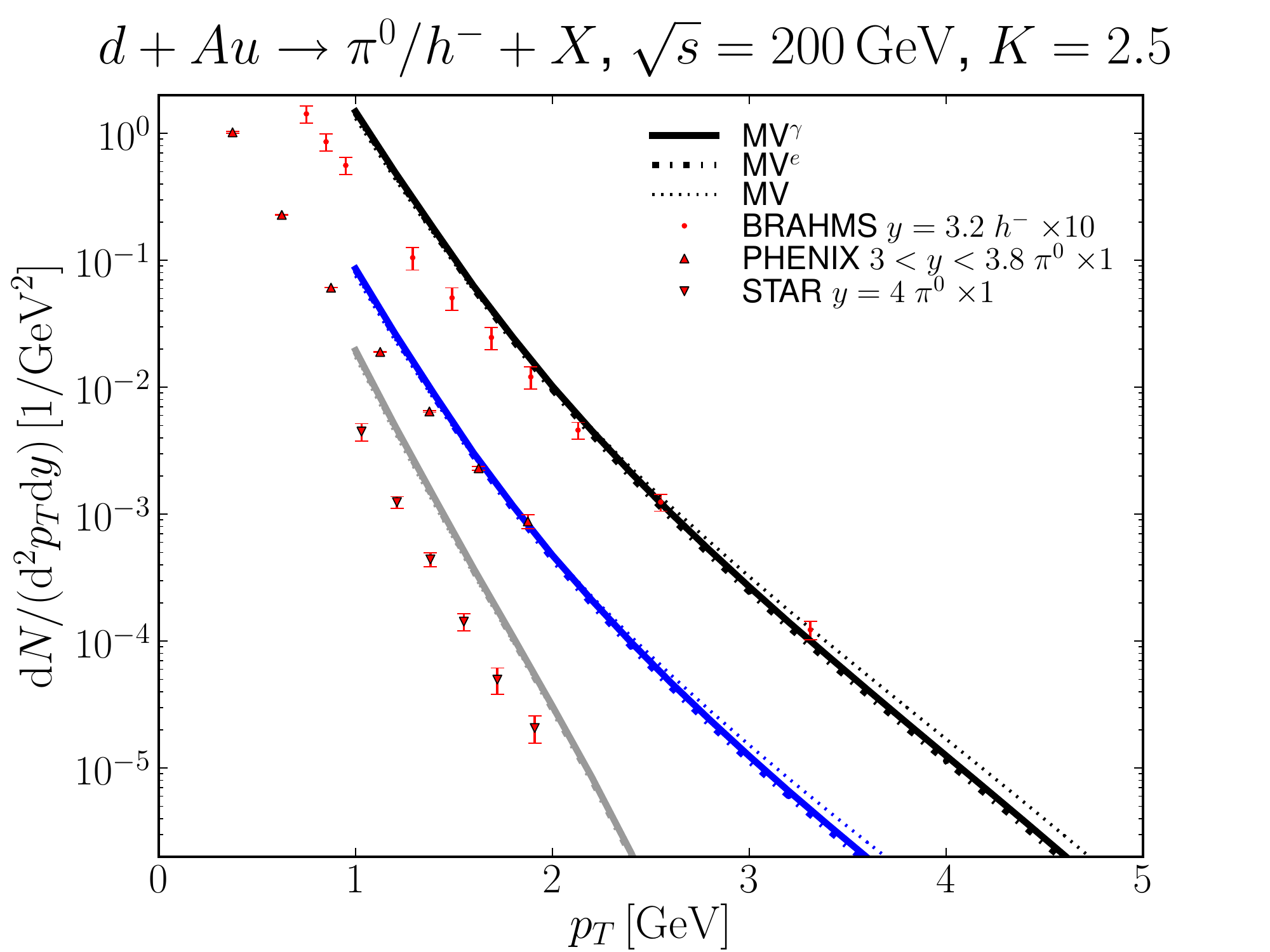}
\caption{
Single inclusive $\pi^0$ and negative hadron production at $\sqrt{s}=200$ GeV d+Au collisions compared with BRAHMS~\cite{Arsene:2004ux}, STAR~\cite{Adams:2006uz} and PHENIX~\cite{Meredith:2011nla} data.\\
}\label{fig:rhic_dau_yield}
\end{minipage}
\hspace{0.5cm}
\begin{minipage}[t]{0.48\linewidth}
\centering
\includegraphics[width=1.05\textwidth]{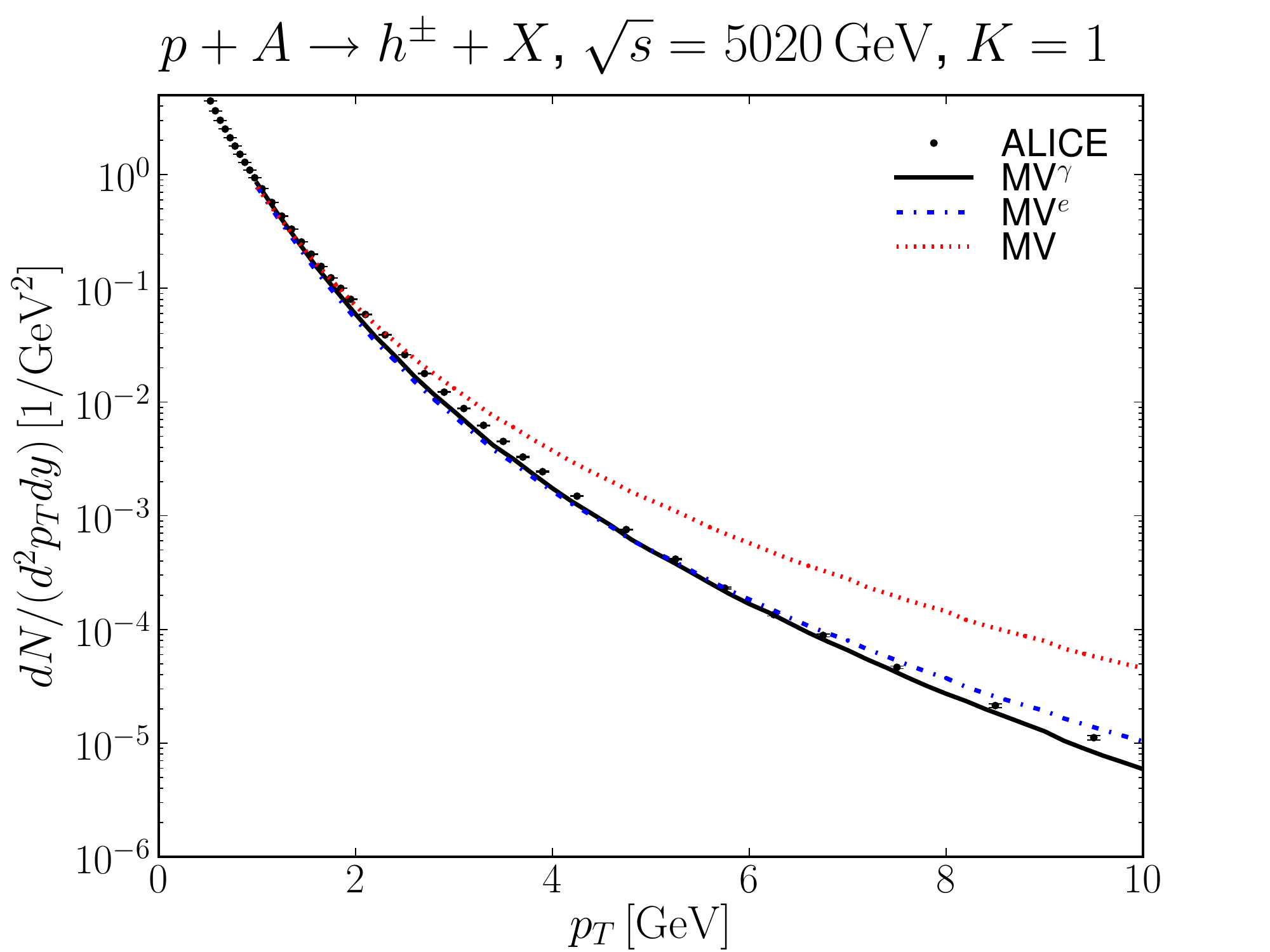}
\caption{
Single inclusive charged hadron production in minimum bias p+Pb collisions at $\sqrt{s}=5020$ GeV computed using the $k_T$ factorization and compared with ALICE data \cite{ALICE:2012mj}.\\
}\label{fig:alice_pa_yield}
\end{minipage}
\end{figure}

\begin{figure}[tb]
\begin{minipage}[t]{0.48\linewidth}
\centering
\includegraphics[width=1.05\textwidth]{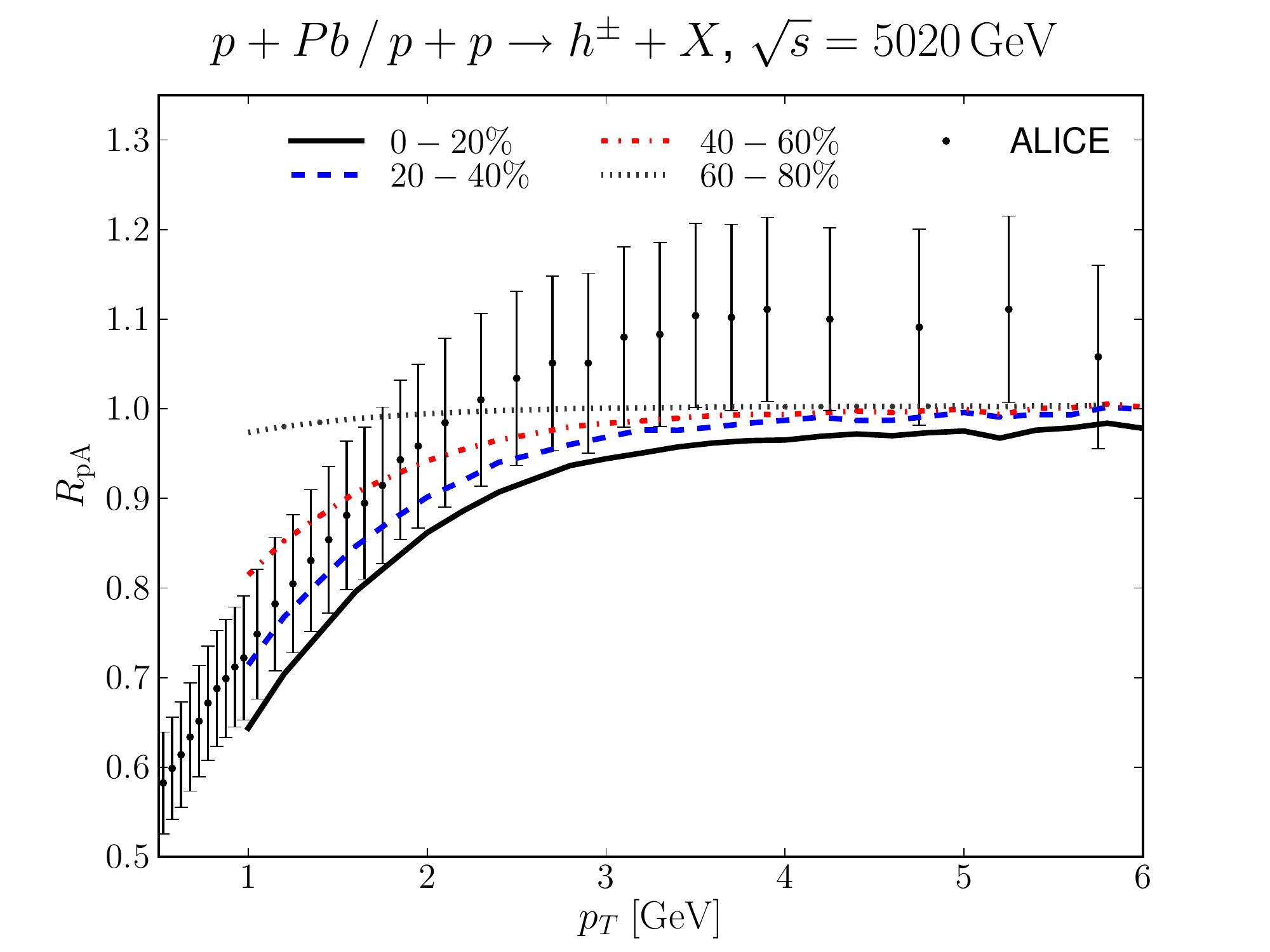}
\caption{
Centrality dependence of the nuclear modification factor $R_{pA}$ at $\sqrt{s}=5020$ GeV p+Pb collisions compared with the minimum bias ALICE data~\cite{ALICE:2012mj}.
}\label{fig:rpa_y0_c}
\end{minipage}
\hspace{0.5cm}
\begin{minipage}[t]{0.48\linewidth}
\centering
\includegraphics[width=1.05\textwidth]{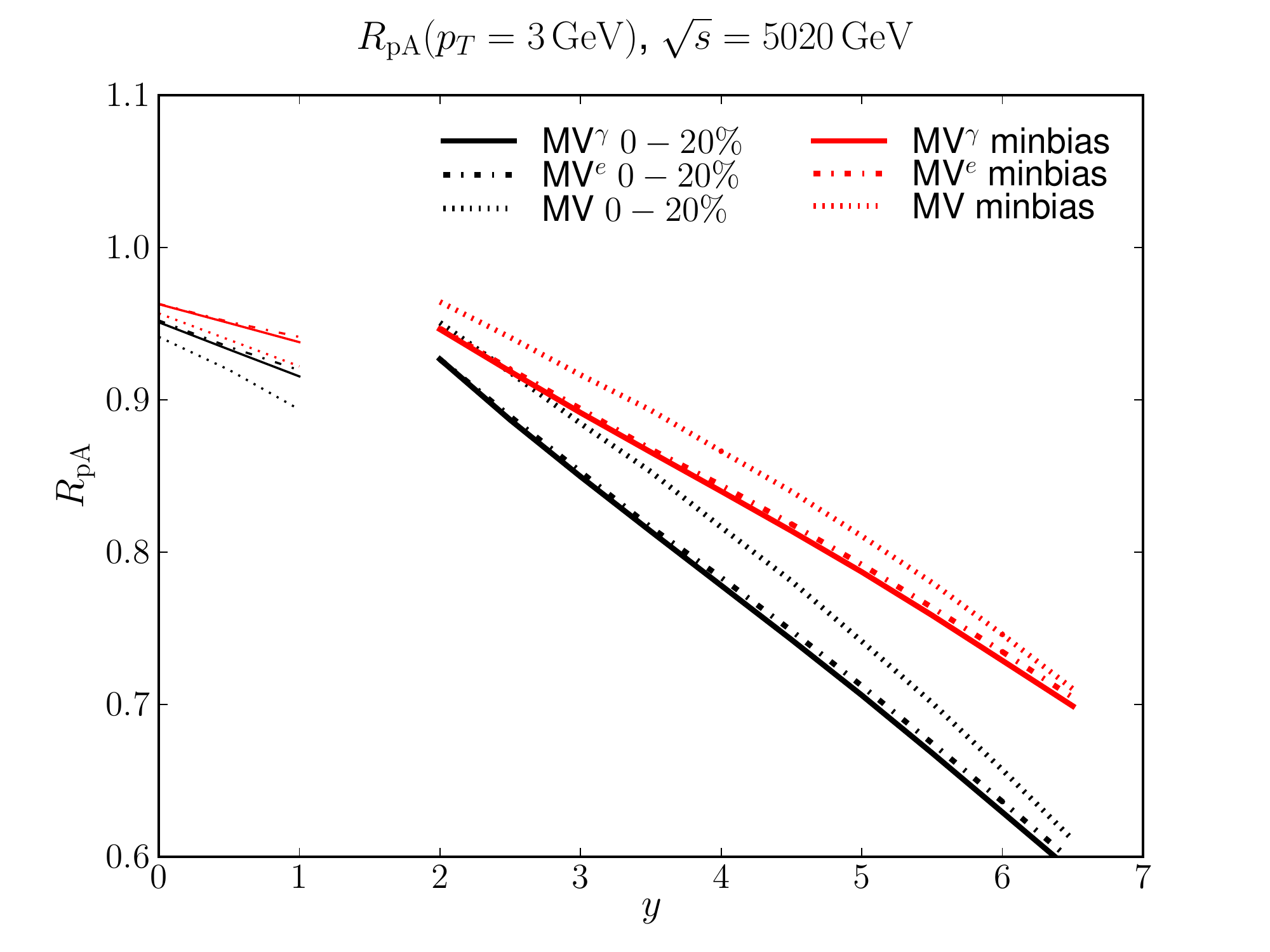}
\caption{
Rapidity dependence of the nuclear modification factor $R_{pA}$ for $3$ GeV neutral pion production at most central and minimum bias collisions.
}\label{fig:rpa_fixedpt}
\end{minipage}
\end{figure}

In Fig. \ref{fig:rhic_pp_yield} we show the single inclusive $\pi^0$ and negative hadron yields computed using the hybrid formalism at $\sqrt{s}=200$ GeV and compare with the experimental data from RHIC \cite{Adams:2006uz,Adare:2011sc,Arsene:2004ux}. As an initial condition for the BK evolution we use MV, MV$^\gamma$ and MV$^e$ fits. We recall that all fits, especially MV$^\gamma$ and MV$^e$, give a good description of the HERA DIS data (see Fig. \ref{fig:sigmar}). We observe that all initial conditions yield very similar particle spectra, and especially the STAR $\pi^0$ spectra work very well, using 
$K=2.5$. The agreement with BRAHMS and PHENIX data is still reasonably good even though the $p_T$ slope is not exactly correct. 


The midrapidity single inclusive $\pi^0$ and charged hadron yields at $\sqrt{s}=7000$ GeV computed using $k_T$-factorization and compared with ALICE~\cite{Abelev:2012cn}  and CMS~\cite{Khachatryan:2010us} data are shown in Fig. \ref{fig:lhc_pp_yield}. Both MV$^\gamma$ and MV$^e$ models describe the data well without any additional $K$ factor. The pure MV model gives a too hard spectrum.


We have also tested that evaluating the parton distribution function or the fragmentation function at NLO instead of LO accuracy, using a different fragmentation function\footnote{See also Ref.~\cite{d'Enterria:2013vba} where it is shown that the NLO pQCD calculations tend to overpredict the LHC and Tevatron spectra at large momenta due to the too hard gluon-to-hadron fragmentation functions.} 
or using the hybrid formalism instead of the $k_T$-factorization does not significantly affect the $p_T$ slope of the single inclusive spectrum. On the other hand the absolute normalization depends quite strongly (up to a factor $\sim 4$) on these choices.

Let us then discuss proton-nucleus and deuteron-nucleus collisions. In Fig. \ref{fig:rhic_dau_yield} we present the single inclusive $\pi^0$ and negative charged hadron production yields in minimum bias deuteron-nucleus collisions at forward rapidities 
computed using the hybrid formalism 
and compared with the RHIC data~\cite{Arsene:2004ux,Adams:2006uz,Meredith:2011nla}. We use here the same $K$ factor $K=2.5$ that was required to obtain correct normalization with the RHIC pp data. The $p_T$ slopes agree roughly with the data, whereas the absolute normalization differs slightly.
In Fig. \ref{fig:alice_pa_yield} we show the single inclusive charged hadron yield in proton-lead collisions compared with the ALICE data~\cite{ALICE:2012mj}. The conclusion is very similar as for proton-proton collisions (see Fig. \ref{fig:lhc_pp_yield}): the pure MV model gives a  wrong $p_T$ slope when comparing with the LHC data, but both MV$^\gamma$ and MV$^e$ models describe the data well.

The centrality dependence of the midrapidity $R_{pA}$ is shown in Fig.~\ref{fig:rpa_y0_c}. Here we only compute results using the MV$^e$ initial condition as $R_{pA}$ depends weakly on the initial condition. The centrality dependence is relatively weak, the results start to differ only at most peripheral classes. Notice that in our calculation we set explicitly $R_{pA}=1$ at centralities $\gtrsim 70\%$, see discussion in Sec. \ref{sec:nuke}. The minimum bias result, which is obtained by averaging over the centrality classes, agrees with the ALICE data.



In order to further demonstrate the evolution speed of the nuclear modification factor we plot $R_{pA}(\pt=3\,\mathrm{GeV})$ for neutral pion production at LHC energies in Fig. \ref{fig:rpa_fixedpt} in central and minimum bias collisions. We compute $R_{pA}$ close to midrapidity using $k_T$-factorization and at forward rapidities using the hybrid formalism.
Thus the obtained curve is not exactly continuous. The evolution speed close to midrapidity (where $k_T$-factorization should be valid) is slightly slower than at more forward rapidities where the hybrid formalism is more reliable.
The MV model initial condition gives a slightly different result than the MV$^\gamma$ and MV$^e$ models, and all dipole models give basically the same evolution speed. Thus $R_{pA}$ is not sensitive to the details of the initial dipole amplitude, and the evolution speed of $R_{pA}$ is driven by the BK evolution. The centrality and especially rapidity evolution speed is significantly faster than in an NLO pQCD calculation using the EPS09s nuclear parton distribution functions~\cite{Helenius:2012wd,helenius2013rapidity}.


\section{Conclusions}
\label{sec:conclusions}
Taking only input from electron-proton deep inelastic scattering and standard nuclear geometry we compute single inclusive hadron production in proton-proton and proton-nucleus collisions from the Color Glass Condensate framework. We find that the MV model initial condition must be modified in order to obtain good description of the LHC single inclusive spectra. We have shown that instead of introducing an anomalous dimension one can also take the infrared cutoff in the MV model to be a fit parameter.


We obtain a good description of the available proton-nucleus and deuteron-nucleus data and get exactly $R_{pA}\to 1$ at large $\pt$ which is a natural requirement and in agreement with the available ALICE data, and follows directly from our consistent treatment of the difference between the proton transverse areas. We present predictions for future forward $R_{pA}$ measurements. Especially we find that the rapidity evolution of the $R_{pA}$ at fixed $\pt$ is a generic prediction of the CGC, given by the BK equation.

\section*{Acknowledgements}
We thank I. Helenius and L. Korkeala for discussions and M. Chiu for providing
us the PHENIX $\pi^0$ yield in proton-proton collisions.
This work has been supported by the Academy of Finland, projects 133005, 
267321, 273464 and by computing resources from CSC -- IT Center for 
Science in Espoo, Finland. H.M. is supported by the Graduate School of 
Particle and Nuclear Physics.

\bibliography{../../refs}
\bibliographystyle{h-physrev4mod2}

\end{document}